\documentclass[amsmat,amssymb,amsfonts,aps,prb,twocolumn,showpacs]{revtex4}

\usepackage{graphicx}
\usepackage{dcolumn}
\usepackage{bm}
\begin{document}

\title{ Hidden multiferroic order in  graphene zigzag ribbons}

\author{J. Fern\'andez-Rossier
%\footnote{jfrossier@ua.es}
}
\affiliation{Departamento de F\'{\i}sica Aplicada, Universidad de Alicante,
Spain }

\date{\today} 

\begin{abstract} 

The insulating magnetic phase in graphene zigzag ribbons, predicted
both by  density functional and mean field Hubbard model calculations,
is described without additional approximations 
with a BCS wave function of two phase-locked condensates 
of spin-polarized electron-hole pairs. The associated  
 order
parameter   is the spin dipole operator that features both 
magnetic and electric  order 
and accounts for the spin-resolved ferroelectricity
of the system. 
Each condensate is associated to
a spin-dependent dipole and their relative phase locking 
sets the total electric dipole and total magnetization equal to zero. 
 
\end{abstract}

\maketitle

%\section{Introduction}

The quest of novel electronic phases, characterized by new
order parameters, and the interplay between electric and magnetic degrees of
freedom are two  of the major themes in condensed matter physics. 
The coexistence of magnetic and electric order
is  associated to transition metal perovskites \cite{multiferroic} 
whereas spintronics proposals are based on materials
where either spin-orbit interactions\cite{Datta} or 
$d$ electrons\cite{Ohno} play a prominent
role.  Here I show that
the magnetic ground state predicted\cite{Fujita96,Waka98,Waka03,zz1,zz2,zz3,zz4} 
in graphene zigzag ribbons,
a chemically simple system without $d$ electrons and negligible spin orbit
coupling, is indeed  a new
electronic phase whose order parameter is the product of the spin and the
electric polarization. 
This new electronic state may be studied experimentally thanks
to recent progress in the fabrication of 
 graphene\cite{Geim05,Kim05,Natmat}
%a single atomic plane of graphite, 
and graphene based flat nanostrucutres \cite{Natmat,Melinda,Pablo07}.

Early theory work predicts that graphene is a zero gap semiconductor, with
electron-hole symmetry and linear conduction and valence bands. These features
arise naturally from a tight-binding model  with one $\Pi_z$ orbital per atom in a
honeycomb lattice at half filling  and they are related to fact that 
the honeycomb
lattice is  bipartite.
% lattice made of two interpenetrating triangular sub-lattices.  
The electronic structure of graphene  nanoribbons depends
dramatically on their atomic strucuture\cite{Nakada96,Brey06,Fede06,JFR07a}.
 Here I focus on graphene
ribbons with zigzag edges. The single-particle description of this system
features  two almost degenerate quasi-flat   bands at the Fermi
energy\cite{Nakada96,Brey06,Fede06}. These flat bands are associated to edge
states . When Coulomb repulsion is added to this picture within  a mean field
Hubbard model\cite{Fujita96,Waka98,Waka03} local moments of oppposite signs  form in
the edges, with a total zero spin, and a gap opens at the Fermi energy. 
The predictions of this model are robust with respect to the addition of more
orbitals, second neighbour hoppings and long range Coulomb interactions, all
present in density functional (DFT) calculations\cite{zz1,zz2,zz3,zz4}.
DFT results and the mean field Hubbard model yield very
similar results for the low energy sector of the electronic structure both
for  zigzag graphene ribbons\cite{Gunlycke}  and nanoislands\cite{JFR07b}. 
This  permits a significant computational simplification
as well as conceptual advantage through 
the use of exact results valid for the Hubbard model \cite{Lieb89}.

In this paper three things are done. First, I show that
the  mean field wave function  of the Hubbard model for 
graphene ribbons with zigzag edges is  that of two phase
locked  BCS
condensates of spin-polarized electron-hole pairs living in the edge bands,
 the only bands affected by the interactions.  Second, 
the BCS electron-hole coherence implicit in the wave funcion is
associated to the existence of spin-resolved 
transverse electric polarizations that yield a zero total electric dipole and
spin when summed. Therefore,  the standard mean field magnetic phase of zigzag
graphene ribbons\cite{Fujita96,Waka98,Waka03,zz1,zz2,zz3,zz4} is
an excitonic insulator phase with a hidden ferroelectric order.  Third, 
the mean field bands are written in terms of the BCS gap and diagonal
self-energies.

Zigzag graphene ribbons are described  
 with a single-orbital tight-binding model
\cite{Nakada96,Fede06} plus a on-site Hubbard repulsion treated in the mean
field approximation at half filling\cite{Fujita96,Waka98,JFR07b}:
\begin{equation}
{\cal H}= \sum_{\vec{r},\vec{r}',\sigma} t_{\vec{r},\vec{r}'}
c^{\dagger}_{\vec{r}\sigma}c_{\vec{r}'\sigma} +
U\sum_{\vec{r}}n_{\vec{r},\uparrow}\langle n_{\vec{r},\downarrow} \rangle
+n_{\vec{r},\downarrow}\langle n_{\vec{r},\uparrow} \rangle
\label{HMF0}
\end{equation}
where $c^{\dagger}_{\vec{r}\sigma}$ creates an electron at the $\Pi_z$
orbital of atom located at $\vec{r}$ with spin $\sigma$,
$n_{\vec{r},\sigma}=c_{\vec{r}\sigma}^{\dagger}c_{\vec{r}\sigma}$ is the
occupation operator. The first term in the Hamiltonian describes the
first-neighbour hopping ($t=2.5eV$) in the graphene ribbon 
and the second describes 
on-site Coulomb repulsion. I take $U=2$eV.   
The zigzag ribbon is a one dimensional crystal
whose unit cell,  shown in fig. 1a, is repeated along the x direction.
The position $\vec{r}$ is determined by a unit
cell index, $x$ and a intra-cell index $I$.
Notice that the top and botton atoms belong to different sub-lattices. 
In a unit cell there are  with $N_P$ pairs  of $A$ and $B$ atoms,
and the width of the ribbon is $W\simeq \sqrt{3} (N)a$, with $N=2N_P$.  

The spectrum of the self-consistent mean field Hamiltonian for a ribbon
with $N$ atoms per unit cell has $N$ bands per spin channel, half of which are
occupied.  Figure 1b shows the well known  non-interacting $(U=0)$
bands for a ribbon with $N=18$. Solid (dashed) lines represent full
(empty) states.  The two flat bands in the outer region of the Brillouin
zone correspond to states that are localized in the edges of the ribbon
\cite{Nakada96}. As shown in figure 1d, they are not really degenerate
except for $ka=\pm \pi$.  At zero temperature the lower (valence) band
is full and the upper (conduction) band is empty. The operators that
annihilate  an electron in those bands are: 
\begin{eqnarray}
e_{k,\sigma}=C_{k\sigma}\equiv \frac{1}{\sqrt{L}} 
\sum_{x,I}e^{ikx} \psi_{ck}(I) 
c_{x,I,\sigma} \nonumber \\
h^{\dagger}_{k,\sigma}=V_{k\sigma}\equiv 
\frac{1}{\sqrt{L}} \sum_{x,I}e^{ikx} \psi_{vk}(I) 
c_{x,I,\sigma}
\end{eqnarray}
where $\frac{1}{\sqrt{L}}e^{ikx} \psi_{(c,v),k}(I)$ are the Bloch eigenstates
and  $L$ is the length of the ribbon.  
The gap between these bands is proportional to the penetration of the edge
states towards the bulk region \cite{Nakada96}. 

Figure 1c  shows the mean field interacting bands, shifted rigidly by
$-U/2$. A gap opens at the Fermi energy, in agreement with DFT
calculations\cite{zz1,zz2,zz3,zz4}.
The average spin-resolved charges   along a unit cell 
are shown in figures 2a,2c.  Spin up (down) electrons pile at the top
(bottom) edge of the ribbon and leave a charge deficit in the opposite
side, also in agreement with DFT calculations.  Therefore, the
edges have local magnetization with opposite sign. For a given spin,
there is an excess of electrons in one edge that are missing in the
other. The total electronic charge turns out to be the same in all the atoms.

\begin{figure}
[hbt]
\includegraphics[width=3.0in]{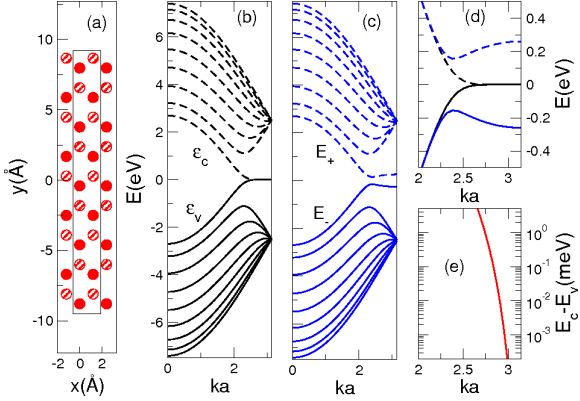}
\caption{ \label{figure1}(Color online).  
(a) Zigzag ribbon unit cell. (b)  Bands of $N=18$ zig-zag ribbon with $U=0$  
(c) and $U=2$eV. Only half of the Brillouin zone is shown.
(d) Comparison of low energy bands. 
Interacting bands have been shift downwards
by $U/2$. (e) $U=0$ single particle gap at the edge states. }
\end{figure}

It is crucial to realize that the non-interacting and the shifted
mean field bands are identical except for
the lowest energy empty
band  and the highest energy occupied band  which differ in the
outer sector of the Brillouin zone, shown in figure 1c. These bands are denoted
by $-$ and $+$ and $v$ and $c$ for the interacting and the non-interacting case.
It turns out that both $-$ and $+$ can be expressed as linear combinations of
$c$ and $v$ {\em only}, with an accuracy better than $99\%$. 
Thus it is possible to
relate the two interacting states $-$ and $+$ with the non-interacting
conduction and valence band through
\begin{eqnarray}
\left(\begin{array}{c} f^{\dagger}_{-k\sigma}  \\ f^{\dagger}_{+k\sigma}
\end{array}\right) =
\left(\begin{array}{cc} u_{k} &  v_{k} e^{i\phi_{\sigma}}\\
- v^*_{k} e^{-i\phi_{\sigma}}&  u^*_{k} \end{array}\right) 
\left(\begin{array}{c} V^{\dagger}_{k\sigma}  \\ C^{\dagger}_{k\sigma}
\end{array}\right)
\label{rotation} 
\end{eqnarray} 
Importantly,  the spin dependence is limited to the phases. 
The moduli of the coefficients $|u_k|^2$ and $|v_k|^2$, are
shown in figure 2b.
We find that  $v_{k}^2+u_{k}^2>0.99$.
Using this relation we formulate the interacting theory in terms of
electrons and holes in the non-interacting valence and conduction
band.  The mean field ground state, which  is formed by filling all the
mean-field bands below the gap, is written as 
$|\Phi\rangle=|\Phi\rangle_{\downarrow}\times|\Phi\rangle_{\downarrow}$ where:
%\begin{equation}
$|\Phi\rangle_{\sigma}=\Pi_{k} f^{\dagger}_{-k,\sigma} |G'\rangle$,
%\end{equation}
where $|G'\rangle$ denotes the state where all the bands below $V$ (or $-$)
 in figures 1b and 1c
are full. Making use of eq. (\ref{rotation}) I write:
\begin{equation}
|\Phi\rangle= \prod_{k,\sigma} \left(u_{k} + v_{k} e^{i\phi_{\sigma}}
e^{\dagger}_{k,\sigma}h^{\dagger}_{k,\sigma}\right)
%\left(u_{k,\downarrow} + v_{k,\downarrow}
%e^{\dagger}_{k,\downarrow}h^{\dagger}_{k,\downarrow}\right)
|G\rangle_{0}
\label{BCS-wave}
\end{equation}
where $|G\rangle_{0}$ %=\Pi_{k\sigma}d^{\dagger}_{k\sigma}|G'\rangle$ 
is the non-interacting
ground state  with no holes in the valence band and no electrons in the
conduction band. Equation (\ref{BCS-wave}) is one of the important results of
this work: the mean field ground state implicit in eq.(\ref{HMF0}) that yields
the bands of fig 1c and the density profile of fig. 2c and 2d 
can be written as  the product of two  BCS condensates of spin polarized
electron-hole pairs. 
This wave function is found in the context of
excitonic insulator \cite{Kohn}
and non-equilibrium exciton condensates  \cite{JFR97}.
%The wave function (\ref{BCS-wave}) portraits the transition from fig. 1b to 1c
%as an excitonic 
%insulator instability. 

\begin{figure}
[t]
\includegraphics[width=3.0in]{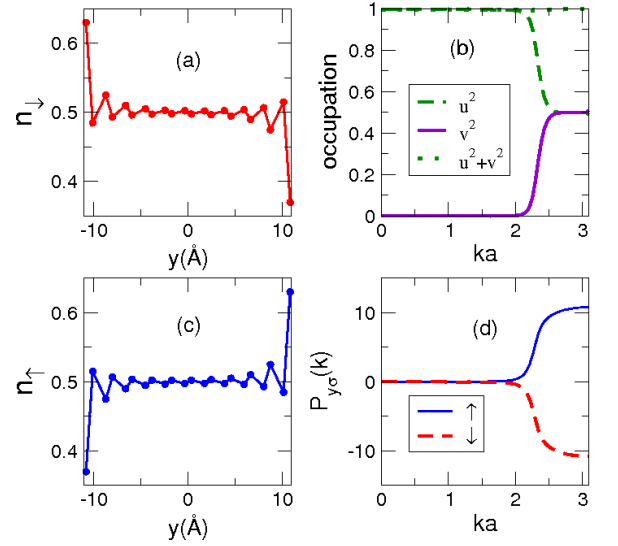}
\caption{ \label{figure2}(Color online).  
 (a) and (c): spin revolved occupation
 $n_{I\sigma}$ as a function of vertical
position in unit cell for $N=22$ ribbon. 
(b): Occupation factors $v_{\sigma}(k)^2$, 
$u_{\sigma}(k)^2$ and $v_{\sigma}(k)^2+u_{\sigma}(k)^2$.  
(d) Spin resolved dipole  ${\cal P}_{y \sigma}(k)$  }
\end{figure}

Importantly, this BCS state
implies the existence of non-magnetic long-range order for the interband 
operators. The numerical
calculations systematically show that the interband coherence 
\begin{equation}
\langle \Phi|e^{\dagger}_{k\uparrow}h^{\dagger}_{k\uparrow}|\Phi\rangle= 
  v^*_{k}u_{k,}e^{-i\phi_{\uparrow}}=
%  -v^*_{k}u_{k}e^{-i\phi_{\downarrow}}
-  \langle \Phi|e^{\dagger}_{k\downarrow}h^{\dagger}_{k\downarrow}|\Phi\rangle
\label{phaselock}
\end{equation}
are finite
and their relative phase is locked: $(\phi_{\uparrow}-\phi_{\downarrow})=\pi$.
Interband coherence is zero for $U=0$ and is related to  
observables that mix the valence and conduction band. These
 acquire an anomalous
expectation value in the $U>0$ phase. The
 interband coherence is associated to electric polarization \cite{JFR98}
in non-equilibrium exciton
condensates  and to electronic ferroelectricity in
the case of Bose-condensation of  slave bosons in the case of mixed-valence
compounds \cite{Sham-Arovas}.
Hence,  I
 look for the connection between the interband coherence  implicit in eq.
(\ref{BCS-wave}) and the spin-resolved electric dipole implicit in figs. 2a,2c.
The electric dipole is written as the sum of the spin-resolved dipole
%\begin{equation}
${\cal P}_y= {\cal P}_{y\uparrow}+{\cal P}_{y\downarrow}$
%\end{equation}
where 
\begin{equation}
{\cal P}_{y\sigma}= \sum_{x,I} y_I en_{x,I,\sigma}= 
\sum_{k,\nu,\nu'}d_{\nu,\nu'}
C^{\dagger}_{k,\nu,\sigma}C_{k,\nu',\sigma}
\end{equation}
are the spin-resolved components of the dipole operator, 
and the dipole matrix elements are given by
\begin{eqnarray}
d_{\nu,\nu'}(k)=
\sum_{I}ey_I\psi_{k,\nu}^*(I)\psi_{k,\nu'}(I)
\label{dip-def2}
\end{eqnarray}
The labels $\nu,\nu'$ run over the non-interacting bands.
The ribbon is centered at $y=0$.
Because of the mirror symmetry of the zigzag unit cell,
 $d_{\nu\nu}(k)=0$, 
so that only the band-mixing terms on eq. (\ref{dip-def2}) can yield a
contribution. The average
dipole operator in the state (\ref{BCS-wave}) is:
\begin{eqnarray}
\langle {\cal P}_y\rangle=\frac{1}{L}
\sum_{k,\sigma} d_{CV}(k) u_{k}v_{k}^*e^{-i\phi_{\sigma}} +
h.c. = \frac{1}{L}\sum_{k,\sigma} {\cal P}_{y\sigma}(k)
\end{eqnarray}
Whereas $\langle {\cal P}_y\rangle=0$
the spin resoved components ${\cal P}_{y\sigma}(k)$, shown in fig. 2d
are finite and with opposite sign. 
A zero net dipole resulting for the sum of two opposite spin-resolved
dipoles is  expected from inspection of figures
2a and 2c and the homogeneous  spin-summed charge distribution.
Thus, the spin resolved dipoles are related to the interband
coherence and the absence of net  electric dipole is related their 
phase locking in eq.(\ref{phaselock}).
In order to characterize this new kind of electronic order,
I introduce the spin dipole
operator:
\begin{equation}
{\cal P}_{\sigma,\sigma'}(y,z)= e\sum_{I} y_I n_{I\sigma}
 S^{z}_{\sigma,\sigma'}=
 \left(\begin{array}{cc} {\cal P}_{y\uparrow} & 0\\
0 & -{\cal P}_{y\downarrow} \end{array}\right)
\label{spin-dipole}
\end{equation}
where $S^{z}_{\eta,\eta'}$ is the Pauli matrix.
Thus, the relevant order parameter associated to
 the electronic state
(\ref{BCS-wave}) is $Tr_{\sigma}\langle \Phi |{\cal P}(y,z)|\Phi\rangle$.
 Spin rotational invariance permits to
choose $z$ along any direction in the spin space. This order parameter is
invariant under the combined action of time reversal and mirror symmetry,
and provides a natural explanation to the spin-polarization of the system when
subject to a transverse electric field, predicted by DFT calculations. Notice
that this phase is different from the non-magnetic ferroelectric phase
predicted in \onlinecite{Waka03}, which  is not found in DFT. 

The mean field  state  (\ref{BCS-wave}) 
invites to write  the interacting bands in terms of a
BCS-like gap related to interband
coherence. To do that, I project out all the bands except $C$ and $V$:
\begin{equation}
c^{\dagger}_{xI,\sigma}\simeq\frac{1}{\sqrt{L}} \sum_{k}e^{-ikx} \left( \psi_{k,C}(I)
C^{\dagger}_{k\sigma} +  \psi_{k,V}(I)
V^{\dagger}_{k\sigma}\right) 
\label{trans}
\end{equation}
The occupation of the sites is expressed as $n_{I\sigma}= \frac{1}{2} +\sigma
m_{I}$, where $\sigma=\pm$. This automatically ensures that the occupation in
each site is 1. 
%This approach is reminecent of the mapping of the electronic structure of
%armchair nanotubes to a two leg Hubbard model \cite{Balents}
Using  transformation eq. (\ref{trans}), the mean field Hamiltonian reads:
\begin{eqnarray}
{\cal H}= \sum_{k,\sigma}
\left(C^{\dagger}_{k,\sigma},V^{\dagger}_{k,\sigma}\right)
\left(\begin{array}{cc}
\xi_{c\sigma}(k) & \Delta_{\sigma}(k) \\
 \Delta^*_{\sigma}(k)& \xi_{v\sigma}(k) 
\end{array}
\right)
\left(\begin{array}{c} C_{k,\sigma}\\V_{k,\sigma}\end{array}\right)
\label{MF-2}
\end{eqnarray}
with 
\begin{eqnarray}
\xi_{\nu\sigma}(k) &=& \epsilon_{\nu}(k)+\frac{U}{2} + 
U\overline{\sigma} \sum_{I} 
|\psi_{k,\nu}(I)|^2
\langle m_I\rangle
\label{selfdiag}
\end{eqnarray}
where $\epsilon_{\nu}(k)$ are the $U=0$ bands and  $\nu=c,v$. The
second term in (\ref{selfdiag}) is the rigid shift 
of the bands $U \frac{n}{2}$ and
the third term is the diagonal self energy $\Sigma_{\nu,\sigma}(k)$.
The off-diagonal self-energy reads:
\begin{eqnarray}
\Delta_{\sigma}(k) &=& \overline{\sigma}U\sum_{I} \psi_{k,c}(I)\psi^*_{k,v}(I)
\langle m_{I}\rangle 
\label{selfgap}
\end{eqnarray}
Notice that $\Delta_{\sigma}(k)=-\Delta_{\overline{\sigma}}(k)$,
 which explains the phase locking of eq. (\ref{phaselock}). 
Notice also that
in the Hubbard model
 the self energies for spin $\sigma$ electrons depend on the density
of carriers with opposite spin $\overline{\sigma}$. 
For each $k\sigma$ the mean field two by two matrix can be written as 
$\frac{U}{2}+\vec{h}_{\sigma}(k)\vec{\tau}$ where $\tau$ are the Pauli matrices,
and the effective field can be written as:
\begin{eqnarray}
\vec{h}_{\sigma}(k)=\left(Re(\Delta_{\sigma}(k)),Im(\Delta_{\sigma}(k)),
\frac{\xi_{c,\sigma}-\xi_{v,\sigma}(k)}{2}\right)
\label{vech}
\end{eqnarray} 
The eigenvalues of this two by two matrix are 
\begin{eqnarray}
E_{\pm,\sigma}(k)=\frac{1}{2}\left(U\pm 
\sqrt{(\xi_{c,\sigma}-\xi_{v,\sigma}(k))^2+4|\Delta_{\sigma}(k)|^2}\right)
\nonumber
\end{eqnarray}
The transformation (\ref{rotation}) permits to diagonalize (\ref{MF-2}),
obtaining 
%\begin{equation}
${\cal H}= \sum_{k,\sigma,\tau=\pm} 
E_{\tau,\sigma}(k)
 f^{\dagger}_{k\tau\sigma}f_{k\tau\sigma} $.
%\label{MF-3}
%\end{equation}
At zero temperature only the lower branches $E_{-,\sigma}(k)$ are occupied.
The 
mean field dispersion $E_{\tau\sigma}(k)$ depends on  $\xi_{\nu\sigma}(k)$
and on $\Delta_{\sigma}(k)$ which in turn depend
on the magnetization:
\begin{eqnarray}
m(I)=
\sum_k
\psi_{k,c}^*(I)\psi_{k,v}(I)
u_{k}v^*_{k} +
h.c.
%\nonumber
\label{mag}
\end{eqnarray}
The magnetization depends on the transformation factors, $u$ and
$v$, which depend on the energies through:
\begin{equation}
|v_k|^2=\frac{1}{2}\left(1-\frac{h_{z\sigma}(k)}{|\vec{h}_{\sigma}(k)|}
\right)\; \;u_kv_ke^{i\phi_{\sigma}}=\frac{-h_{x,\sigma}(k)}{|\vec{h}_{\sigma}(k)|} 
\label{vself}
\end{equation}
% where I have omitted the $k,\sigma$ indexes.
Equations
(\ref{selfdiag},\ref{selfgap},\ref{vech},\ref{mag},\ref{vself}) form
a self-consistent set.  The numerical
solutions of the mean field Hubbard model (\ref{HMF0}) also satisfy
these equations. This permits to relate the mean field dispersion
to the diagonal self energy $|\Sigma_{c,\sigma}(k)-\Sigma_{v,\sigma}(k)|$
and the off diagonal self energy $|\Delta(k)|$, shown in fig. 3. 
They  are spin indepedent.  It is apparent
that these self-energies are finite in different regions of the Brilloiun zone.
The diagonal self-energy are
related to the non hybridized edge states located in the outer region
of the Brillouin zone whereas the off-diagonal self-energy occurs for
weakly hybridized edge states, at smaller $|k|$.
Further reduction of  $|k|$ opens the  
single-particle gap, which  overshades the self-energies.
These results also permit to unveil the origin of the gaps
$\Delta^1$
 and $\Delta^0$ introduced in \onlinecite{zz2} (see  fig. 3). 
It is apparent that  
the $\Delta^1$ gap is given by  the diagonal self-energy whereas the
$\Delta^0$ gap is related to the non-diagonal self-energy. 
Accordingly, $\Delta^1$ is insensitive to the ribbon width
and can be  approximated by $ \Delta^1\simeq 2 U |m|$
where $|m|$ is the magnetization of the edge atoms.
In contrast,   
$\Delta^0$ decreases as 
 the ribbon width increases due to the smaller 
 inter-edge hybridization.

\begin{figure}
[t]
\includegraphics[width=2.8in]{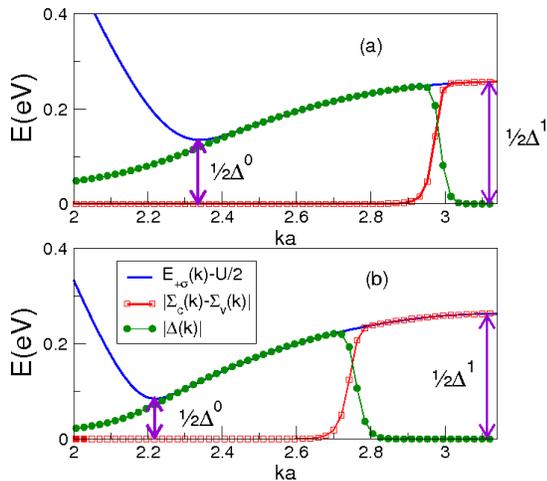}
\caption{ \label{figure2}(Color online).  
Relation between dispersion $E_{+\sigma}(k)$ and 
the diagonal $\Sigma_{\nu\sigma}(k)$
 and off-diagonal $\Delta(k)$ self-energies for two  ribbons with $N=22$ (a)
and $N=42$ (b).}
\end{figure}

The long range order implicit in this and previous mean field theories of
graphene zigzag ribbons  \cite{Fujita96,Waka98,zz1,zz2,zz3,zz4} is known to be
destroyed in one dimension because of long-wavelength  spin wave modes\cite{Waka98}
 associated to the breaking
of a continous symmetry. I have verified that the results of this work remain
valid for finite length graphene ribbons and tubes for which Goldstone modes
have a confinement gap. Therefore, the results of infinite ribbon systems are
relevant for finite systems. 

%The
%coupling of spin waves to electric polarization modes might also result in
%long-range interactions that open a Higgs gap. The mean field approach
%is known to underestimate the smallest $U_c$ that  results in a 
%magnetic inestability in  two dimensional graphene \cite{Sorella}. 
%In graphene zigzag ribbons the  mean field  critical $U$ is zero
%\cite{Fujita96}. 

 In summary, the  BCS wave function (\ref{BCS-wave}) 
 describing two phase-locked
 condesates of spin-polarized electron hole pairs
 is the the collective wave function behind
 the  insulating ferrimagnetic phase
in graphene zigzag ribbons portrayed by mean field Hubbard
\cite{Fujita96,Waka98,Waka03} model, which yields the same results than
 DFT calculations \cite{zz1,zz2,zz3,zz4}.
 The underlying electron-hole coherence in each
spin-channel is  related to mirror symmetry  breaking of the charge
density for a given spin. Their relative phase-locking warrants that the total spin and
electric dipole are zero. The natural order parameter for this  electronic
state with magnetic and spin-hidden electric order 
 is the spin-dipole operator (eq. (\ref{spin-dipole}). The reformulation of
the mean field theory in terms of  a 2-band  BCS model rationalizes the
shape of the mean field bands  in terms of  diagonal and non-diagonal 
self-energies. The joint presence of electric and magnetic order anticipates
non-trivial magnetoelectric effects in graphene zigzag ribbons.

Fruitful converstations with J. J. Palacios and L. Brey are
acknowleged. This work has been financially supported by MEC-Spain (Grants
MAT2007-65487  and Ramon y Cajal Program), by Generalitat Valenciana
(Accomp07-054),  by Consolider CSD2007-0010 and, in part, by   FEDER funds.

\widetext
\end{document}